\documentclass{emulateapj}

\usepackage{amsmath}

\newcommand{\be}{\begin{displaymath}}
\newcommand{\ee}{\end{displaymath}}

\def\lsim{\hbox{\rlap{\raise 0.425ex\hbox{$<$}}\lower 0.65ex\hbox{$\sim$}}}
\def\gsim{\hbox{\rlap{\raise 0.425ex\hbox{$>$}}\lower 0.65ex\hbox{$\sim$}}}

\def\arcdeg{\mbox{$^\circ$}}

\shorttitle{Estimating First-Light and Rise Times of SNe~Ia}
\shortauthors{Zheng et al.}

\begin{document}

\title{An Empirical Fitting Method for Type Ia Supernova Light Curves. II. Estimating the First-Light Time and Rise Time}

\author{WeiKang Zheng\altaffilmark{1,2},
Patrick L. Kelly\altaffilmark{1}, and
Alexei V. Filippenko\altaffilmark{1,3}
}

\altaffiltext{1}{Department of Astronomy, University of California, Berkeley, CA 94720-3411, USA}
\altaffiltext{2}{e-mail: zwk@astro.berkeley.edu}
\altaffiltext{3}{Senior Miller Fellow, Miller Institute for Basic Research in Science, University of California, Berkeley, CA 94720, USA}

\begin{abstract}
We investigate a new empirical fitting method for the optical light curves of Type Ia supernovae
(SNe~Ia) that is able to estimate the first-light time of SNe~Ia,
even when they are not discovered extremely early.
With an improved ability to estimate the time of first light for SNe Ia,
we compute the rise times for a sample of 56 well-observed SNe~Ia.
We find rise times ranging from 10.5 to 20.5 days, with a mean of 16.0 days,
and confirm that the rise time is 
generally correlated with the decline rate $\Delta m_{15}(B)$, but with large scatter.
The rise time could be an additional parameter to help classify SN~Ia subtypes.

\end{abstract}



\section{Introduction}\label{s:intro}

Type~Ia supernovae (SNe~Ia) are believed to be thermonuclear runaway explosions of
carbon/oxygen white dwarfs (see, e.g., Hillebrandt \& Niemeyer 2000 for a review).
The relationship between the light-curve decline and the peak brightness, known as
the ``Phillips relation" (Phillips 1993; Phillips et al. 1999), makes SNe~Ia excellent calibratable candles, most notably leading to the discovery of the accelerating
expansion of the Universe (Riess et al. 1998; Perlmutter et al. 1999).
 
While much effort has been focused on the post-maximum part of SN~Ia
light curves and an important parameter $\Delta m_{15}(B)$
(e.g., Phillips 1993; Phillips et al. 1999),
there are few studies of the pre-maximum rise, in part because very
few SNe~Ia were discovered at extremely early times --- hence, it was
difficult to derive accurate rise times.
But the number of known very young SNe~Ia has increased in the past few years,
making it possible to measure their initial behaviour and estimate their first-light time.\footnote{SNe~Ia may experience a ``dark phase'' (which could last for a few hours to days) between the moment of explosion and the first observed light (e.g., Rabinak et al. 2012; Piro \& Nakar 2013, 2014).}
Riess et al. (1999) used 30 early-time unfiltered CCD observations of SNe~Ia
and adopted the commonly known $t^2$ function (from the expanding fireball model;
see, e.g., Arnett 1982; Riess et al. 1999; Arnett et al. 2016) to measure the rise time,
finding $t_{r} = 19.5 \pm 0.2$ days. 
Conley et al. (2006) used a larger sample, 73 SNe~Ia from the Supernova Legacy Survey and
also adopted a $t^2$ function, deriving a similar rise time of 19.34 days.
Hayden et al. (2010) used a ``2-stretch" fit algorithm which estimates the rise and
fall times independently, determining a shorter rise time of $17.38 \pm 0.17$ for a set
of 391 SNe~Ia from the Sloan Digital Sky Survey-II. This was followed by Ganeshalingam et al. (2011),
who used a similar two-stretch template-fitting method and found that SNe~Ia with high-velocity
spectral features have a shorter rise time ($t_{r} = 16.63 \pm 0.29$ days) than normal
SNe~Ia ($t_{r} = 18.03 \pm 0.24$ days). 

Ganeshalingam et al. (2011) also showed that
the initial rise of a SN~Ia light curve follows a power law ($t^n$) with index $n=2.20$.
Nugent et al. (2011) found a rising index of 2.01 for the light curve of SN~2011fe, 
very consistent with the commonly known $t^2$ function,
and they use this to estimate the object's first-light time.
More SNe~Ia were subsequently discovered at extremely early times, and thus the first-light time
could be derived typically from the power law ($t^n$) function --- e.g., SN~2012cg (Silverman et al. 2012b),
SN~2013dy (Zheng et al. 2013), iPTF13ebh (Hsiao et al. 2015), ASASSN-14lp (Shappee et al. 2015),
and the {\it Kepler} objects KSN 2012a, KSN 2011b, and KSN 2011c (Olling et al. 2015).
Firth et al. (2015) adopted a more general
$t^n$ model to study a sample of 18 SNe~Ia and found a mean uncorrected rise time
of 18.98 days, with $n=1.5$ to $>3.0$ and a mean value of 2.44.  

In this paper, we apply an empirical fitting method proposed by Zheng \& Filippenko (2017)
to a large sample of SNe~Ia with well-observed optical light curves and thereby
estimate the explosion time $t_0$. Once $t_0$ is derived,
the rise time $t_r$ is easily determined as long as the time of peak brightness, $t_p$, 
is also measured.
Note that the quantity we determine from the data is actually the first-light time ($t_{0f}$) rather than the
true explosion time ($t_0$). However, here we do not distinguish between the two values; namely, we assume
$t_{0f} \approx t_0$, and use $t_0$ as the first-light time throughout the paper.



\section{Analysis}
\subsection{Fitting Method}\label{s:FittingMethod}

Zheng \& Filippenko (2017) proposed an empirical function that can well fit the
optical light curves of SNe~Ia. It is a variant of the broken-power-law function 
shown as
\begin{equation}
L = A'\left(\frac{t-t_0}{t_b}\right)^{\alpha_{r}} \Big{[} 1 +
\left(\frac{t-t_0}{t_b}\right)^{s{\alpha}_{d}}\Big{]}^{-2/s},
\label{eq_lvtbkn2}
\end{equation}
where $A'$ is a scaling constant, $t_0$ is the first-light time, $t_b$ is the break time,
${\alpha}_r$ and ${\alpha}_d$ are the two power-law indices before and after the break
(respectively), and $s$ is a transition parameter. 
This function\footnote{Note that Equation~\ref{eq_lvtbkn2} is mathematically very
similar to the generalised Pareto distribution in statistics (e.g., Abd Elfattab et al. 2007;
Raja \& Mir 2013). It is unclear whether there is any physical connection
between the light-curve shape and the generalised Pareto distribution; further studies are needed to further elucidate possible associations.}
is mathematically analytic, derived directly from the photospheric velocity evolution
function with some reasonable assumptions. 
Zheng \& Filippenko (2017) show that this function can well fit the optical 
light curves of the prototypical Type Ia SN~2011fe, where they fixed $t_0$ during
the fitting process because it is well estimated (see Nugent et al. 2011).
In the following section, we expand this method to a set of nine extremely well-observed 
SNe~Ia, estimating their values of $t_0$ (which means we treat $t_0$ as a free 
parameter during the fitting).
We then apply the method to a larger sample of SNe~Ia having well-observed optical 
light curves.

\subsection{Fitting to the Extremely Good Sample}\label{s:data_analysis1}

We adopt an IDL implementation of {\it mpfit} (Markwardt 2009)
\footnote{https://www.physics.wisc.edu/$^\sim$craigm/idl/fitting.html}
for all the fitting procedures.
During our test fitting, we found that it is difficult to well constrain $t_0$
if $t_0$, $\alpha_r$, and $s$ are all set free, because these three
parameters are related when estimating $t_0$.
Hence, it is necessary to fix a few other parameters in order to get
a good constraint on $t_0$. But before doing that,
it is important to understand each parameter and statistically study them with
at least a few SNe~Ia.

Fortunately, there are nine SNe~Ia that were discovered very early and monitored
well thereafter, making them suitable for our purposes. 
Their first-light times are well estimated directly from the data; thus, for these 
SNe~Ia, we could fix $t_0$ during the fitting and then study the properties of the 
other parameters in Equation~\ref{eq_lvtbkn2}.
This well-observed SN~Ia sample (see Table~1) includes
SN~2009ig (Foley et al. 2009), SN~2011fe (Nugent et al. 2011), SN~2012cg (Silverman et al. 2012b),
SN~2013dy (Zheng et al. 2013), iPTF13ebh (Hsiao et al. 2015), ASASSN-14lp (Shappee et al. 2015),
and three SNe~Ia discovered by the {\it Kepler} spacecraft (KSN 2012a, KSN 2011b, and KSN 2011c; Olling et al. 2015).
Most of them are normal SNe~Ia, except for the three {\it Kepler} SNe with
unknown subtype since we only have broadband light curves
but no spectra, and iPTF13ebh which was categorized
as a ``transitional'' event between normal SNe~Ia and the 
fast-declining subluminous SN~1991bg-like objects 
owing to its relative largely $\Delta m_{15}(B)$ 
(1.79; see Hsiao et al. 2015).

We first fit the light curves by fixing $t_0$ 
because it is already well estimated for each SN.
We then apply the fitting independently to each filter, including the
data points from the beginning of the observational campaign until
the SN entered the phase dominated by cobalt decay, which usually happens around three
weeks after peak brightness in the $B$ and $V$ bands.

Since SNe~Ia generally exhibit a shoulder in the $R$ band and a second peak in the $I$ band,
we restrict the fits to earlier times in the
redder bands ($R$, $I$) than in the bluer bands ($B$, $V$). 
Specifically, for the $R$ and $I$ bands, the data points are cut off 
before the shoulder appears, but still after the main peak.
For a few cases, additional $U$-band data or $g$, $r$, or $i$ filters 
were used. Data with all of these filters are treated independently.
Additionally, the three {\it Kepler} SNe were observed with a broad filter
that is quite different from any existing standard filter. 
However, since we fit all filtered data independently, we apply 
the same fitting to the three {\it Kepler} SNe. Only when plotting, 
we represent these objects with the $R$ band, since the {\it Kepler} 
response function covers the wavelength range 420--900~nm and peaks
around 600~nm (see the {\it Kepler} 
website\footnote{https://keplerscience.arc.nasa.gov/the-kepler-space-telescope.html}).

\begin{figure}
\centering
\includegraphics[width=.5\textwidth]{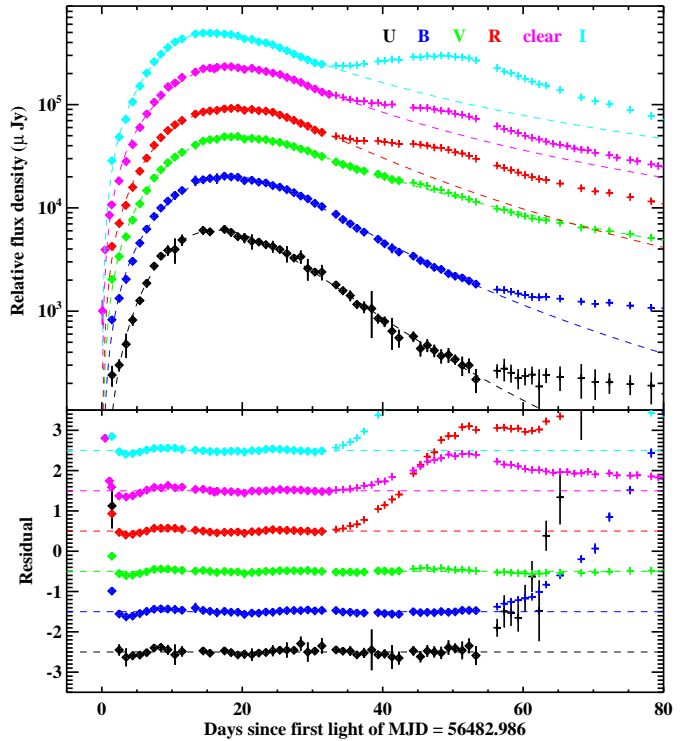}
\caption{Multiband light-curve fitting of SN~2013dy using Equation~\ref{eq_lvtbkn2},
         arbitrarily shifted for clarity.
         The value of $t_0$ is fixed to be $t_{0f}$ (MJD 56482.986)
         during the fitting. Diamond-shaped data points are included in the fitting
         while cross-shaped ones are excluded.}
\label{13dy_IaBkns_fitting_case1}
\end{figure}

The fitting procedure is very similar for each SN as well as for each filter;
thus, we show only one case as an example. Figure~\ref{13dy_IaBkns_fitting_case1}
illustrates the fitting results for SN~2013dy, with comments similar to those for
the fitting of SN~2011fe by Zheng \& Filippenko (2017; see their Figure~2).

As shown in Figure~\ref{13dy_IaBkns_fitting_case1}, by using just one single function
given by Equation~\ref{eq_lvtbkn2}, the fitting results are surprisingly good
for the included data, which cover a long time range and nearly a factor of 100 in flux
($\sim 5$\,mag) for all filters ($U$, $B$, $V$, $R$, $clear$, and $I$).
In particular, for the $V$ band, although we include only the data up to day
45, the model still matches the data well up to nearly three months after the explosion
(though this might be just a coincidence).
The flux residuals are mostly within the $1\sigma$ measurement uncertainties.

Using the same procedure, we apply the fitting to the eight other well-observed
SNe~Ia by fixing $t_0$. In general, the fitting results are very similar to those
of SN~2013dy shown in Figure~\ref{13dy_IaBkns_fitting_case1}, though
they do show some indication for diversity at extremely early times.
For a few cases in our sample including SN~2012cg
and SN~2013dy, we find for the first few days (within three days after first-light time)
that the data show an excess compared to the fit model, possibly from other contributions
such as thermal emission produced by the impact of the SN shock on a binary companion
star (e.g., Kasen 2010; Cao et al. 2015; Marion et al. 2016). In principle, for our purposes
it is more appropriate to exclude these data from the fitting; we therefore
excluded data within 3.0 days after $t_0$ during the fitting.
However, in our later analysis of the first-light time
estimation (in Section \ref{s:data_analysis2}), we find that this early-time excess
insignificantly affects the estimate (the changes are much smaller than the fitting error itself),
so in Section \ref{s:data_analysis2} we retain these data during the fitting process.

\begin{figure*}
\centering
\includegraphics[width=.42\textwidth]{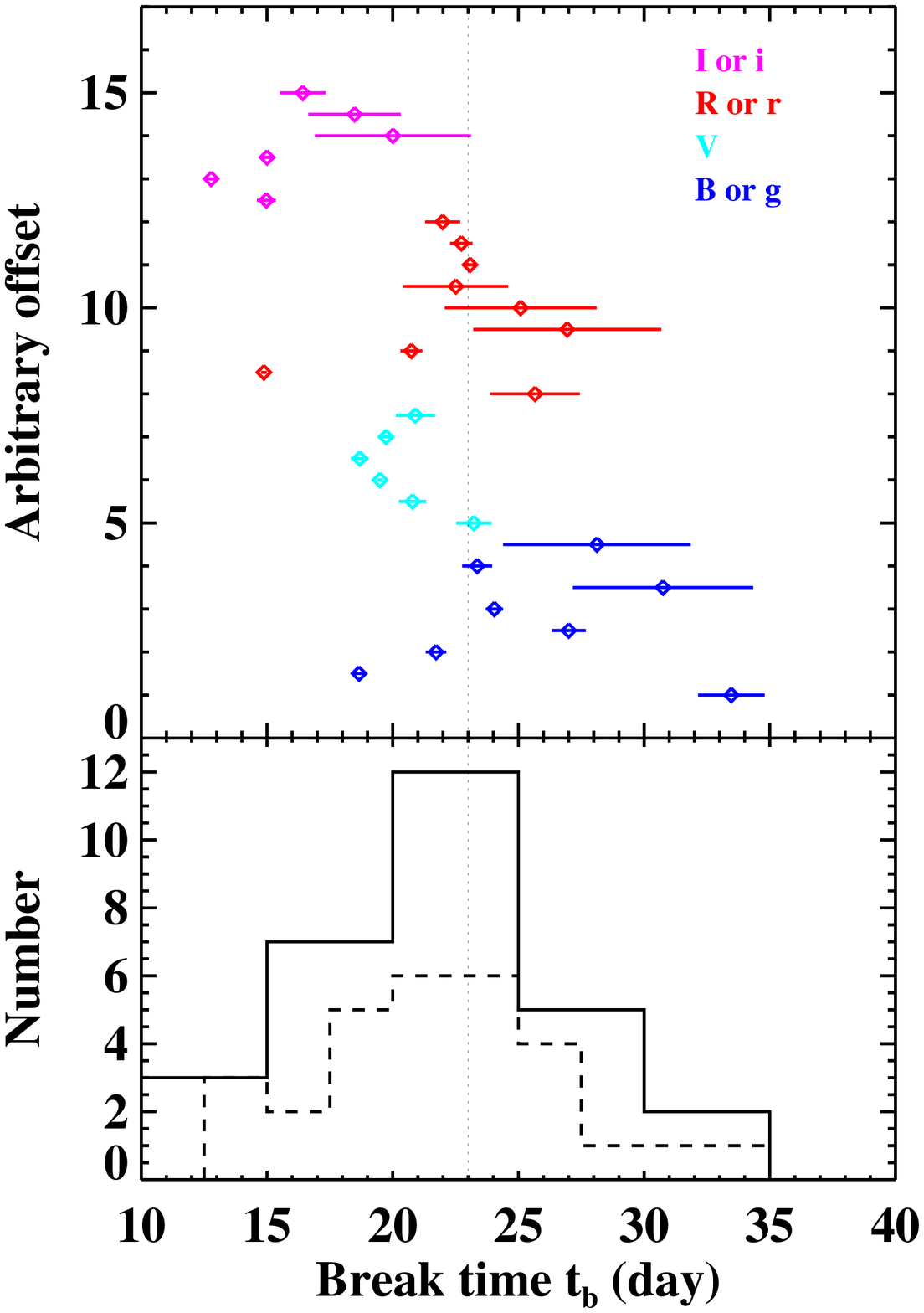}
\includegraphics[width=.42\textwidth]{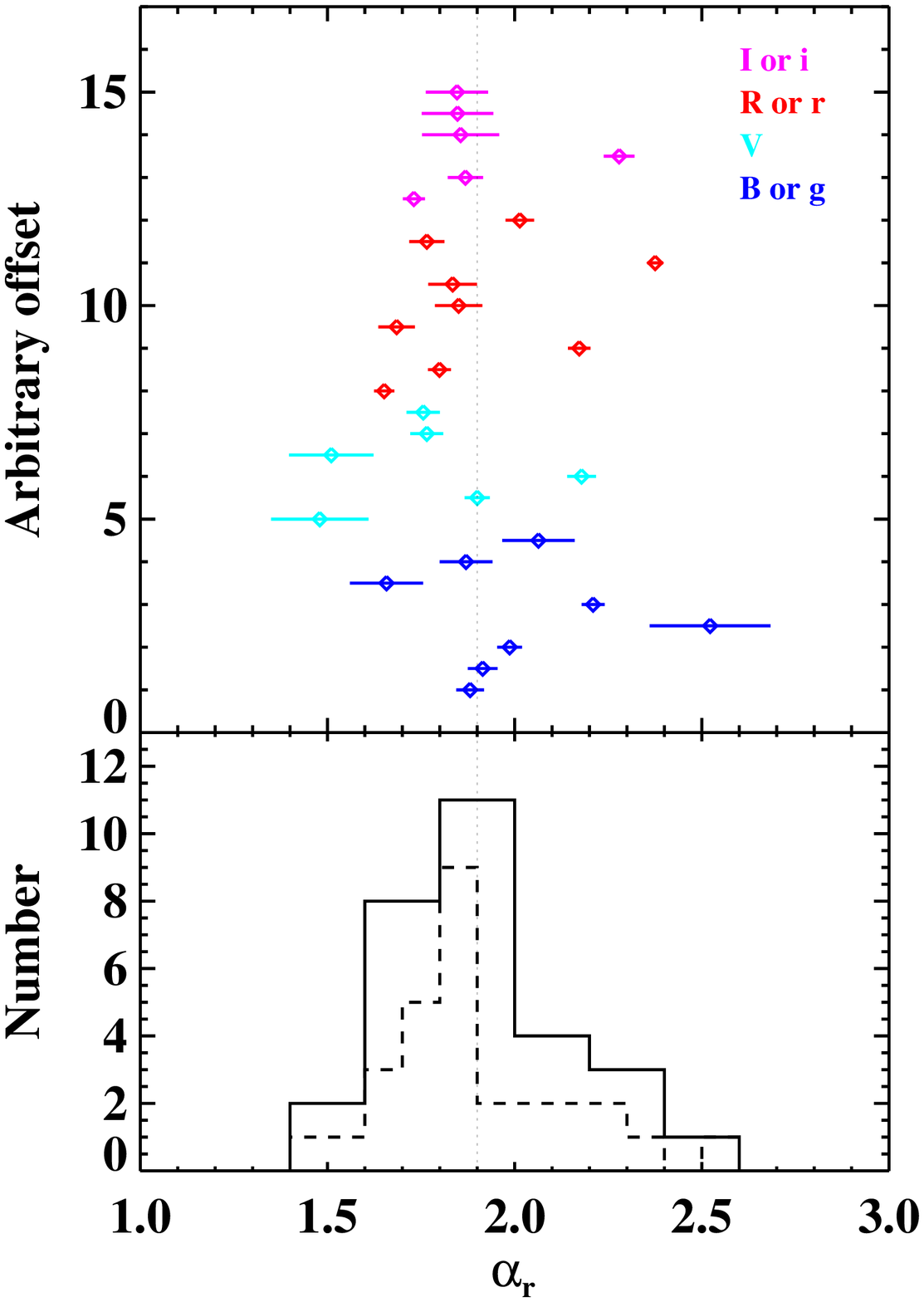}
\includegraphics[width=.42\textwidth]{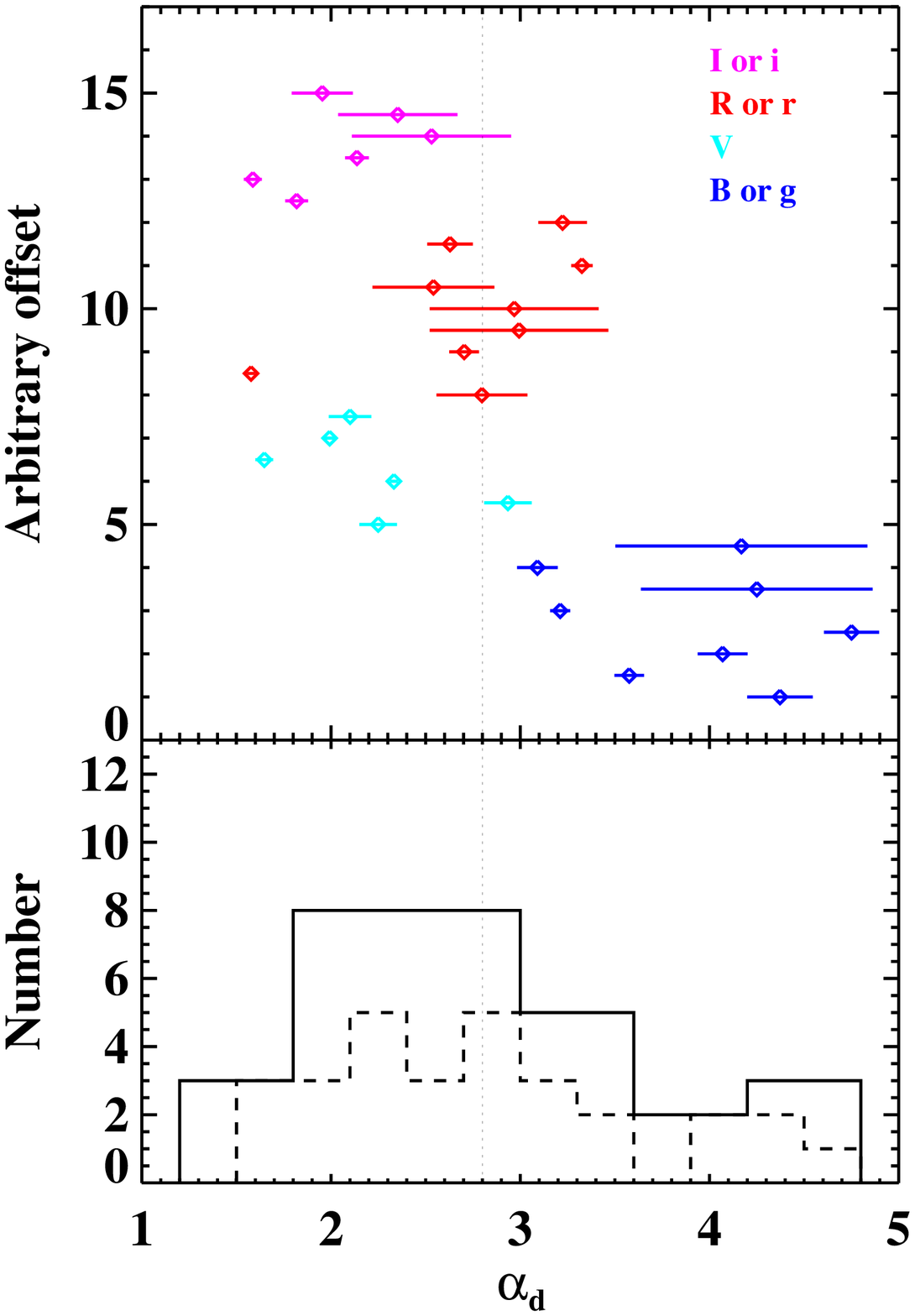}
\includegraphics[width=.42\textwidth]{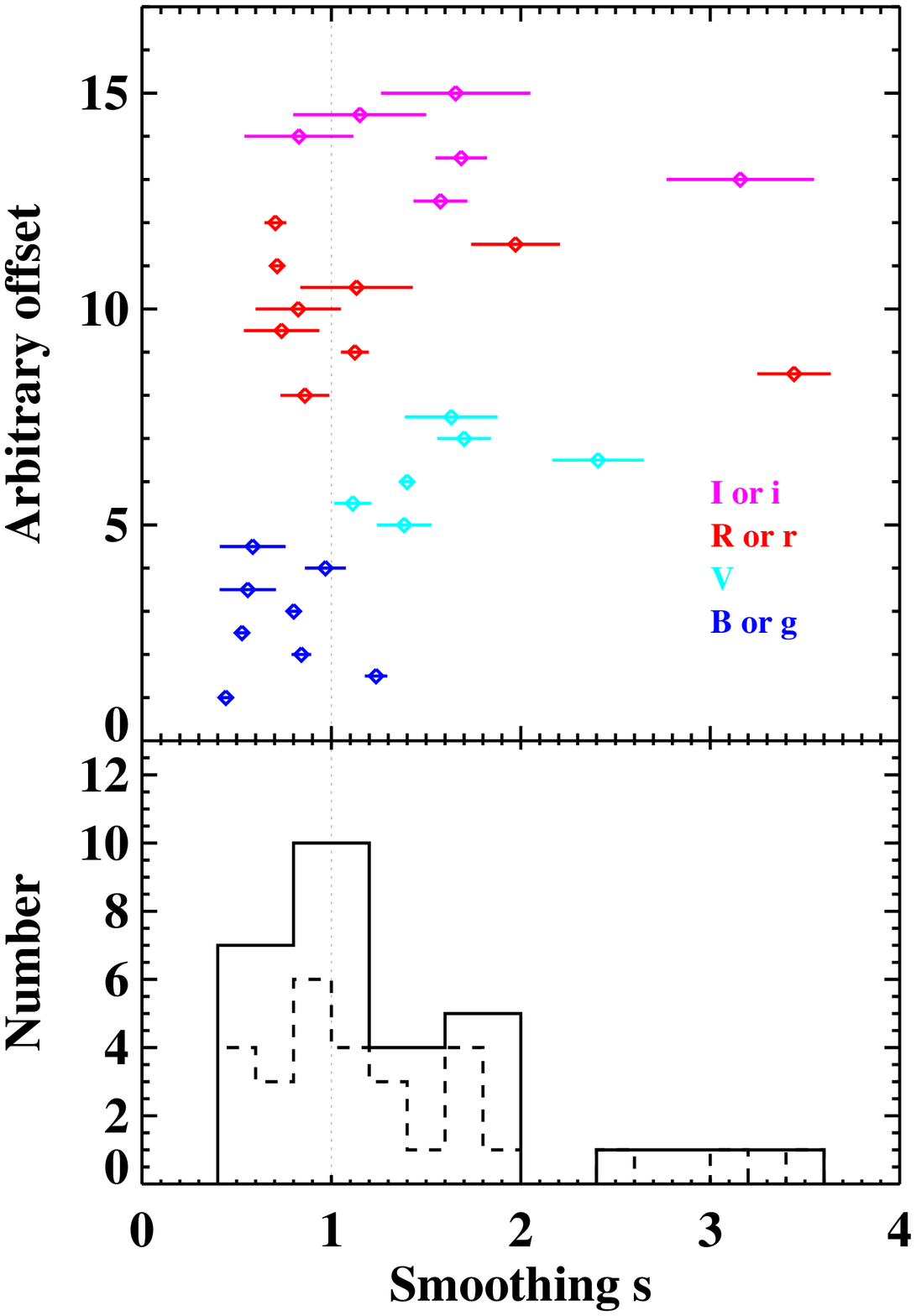}
\caption{Individual points and histogram distribution of each fit parameter from
         Equation~\ref{eq_lvtbkn2}, including $t_b$ (top left), $\alpha_r$ (top right),
	 $\alpha_d$ (bottom left), and $s$ (bottom right). For each parameter, the
	 upper panel shows the individual value from different filters and the
	 lower panel displays the histogram distribution with
	 two different bin sizes.}
\label{fig_Patameters_with_fix_t0}
\end{figure*}

After fitting to all 9 of the SNe given above, we plot the histogram distribution
in Figure~\ref{fig_Patameters_with_fix_t0} for each parameter from Equation~\ref{eq_lvtbkn2},
including $t_b$, $\alpha_r$, $\alpha_d$, and $s$, but not $t_0$ (which was fixed
during the fitting procedure) and $A'$ (which is simply a scaling factor). 
Note that since some SNe lack
observations in one or a few filters, not all filters have the 
complete set of 9 data points representing 9 SNe; however, whichever filter
was observed is fitted independently and shown in the figure.
These fitting results are also listed in Table~1.

As shown in Figure~\ref{fig_Patameters_with_fix_t0}, it is clear that
$\alpha_r$ (top-right panel) is the most concentrated
parameter, with a mean value of $1.90 \pm 0.18$ - very consistent with the commonly known
$t^2$ model for most SNe~Ia or the $t^n$ model ($n$ varies from $\sim1.5$ to $\sim3.0$) studied by
various groups (e.g., Conley et al. 2006; Ganeshalingam et al. 2011; Firth et al. 2015).
The parameter with the next-smallest dispersion is the break time $t_b$ (top-left panel), with a mean value of
$23.0 \pm 4.0$ days, typically a few days after the time of peak brightness.
The other two parameters, $s$ (the smoothing parameter) and $\alpha_d$, are much less concentrated.

In conclusion, to better estimate the first-light time $t_0$
(as shown in the next section), it is appropriate to keep $\alpha_r$ and $t_b$ constant
during the fitting procedure because for each filter these two parameters exhibit a small dispersion among
different SNe. Conversely, it is best to keep $s$ and $\alpha_d$ as free parameters, since they are quite diverse. By doing this, Equation~\ref{eq_lvtbkn2} has fewer
free parameters (only 4, after fixing $\alpha_r$ and $t_b$), and we can better constrain $t_0$ as demonstrated below.

\subsection{Estimating $t_0$ with the Extremely Good Sample}\label{s:data_analysis2}

In reality, very few SNe~Ia have been discovered extremely early like 
the nine SNe mentioned above.
However, for SNe~Ia that were discovered reasonably early (1--2 weeks 
after explosion), one may estimate the explosion time $t_0$ by fitting 
the light curve using Equation~\ref{eq_lvtbkn2}.

Before applying this method to a larger set we need to test its validity.
Our set of 9 well-observed SNe~Ia provides a perfect sample for such test, since all of them
already have a well-determined $t_0$ (which we denote as ``known $t_0$''). We can then compare
the ``known $t_0$ and the $t_0$ value estimated from fitting Equation~\ref{eq_lvtbkn2}
to study the errors from the fitting method.

First, in order to simulate the relatively late-time discovery for the majority of less well-observed SNe~Ia,
we intentionally exclude some of the very early-time data points for the above 9 SNe. However,
a criterion needs to be established for how late to start using the data. Here we define
the limit to be the time prior to peak brightness when the
SN is 1\,mag fainter than the peak magnitude, denoted
by $t_{\rm pm-1}{-0}$; it typically occurs around 1-2 weeks before peak brightness. 
The first data point must be at least 1\,mag fainter than the peak magnitude.
We adopt this limit for selecting SNe to be fitted with Equation~\ref{eq_lvtbkn2}
because we require relatively early-time observations to constrain $t_0$. 
If there are additional observations one day earlier than that, their time will be denoted 
by $t_{\rm pm-1}{-1}$, and so on.
The discovery times of the above 9 SNe are extremely early,
in some cases reaching $t_{\rm pm-1}{-7}$.

\begin{figure}
\centering
\includegraphics[width=.5\textwidth]{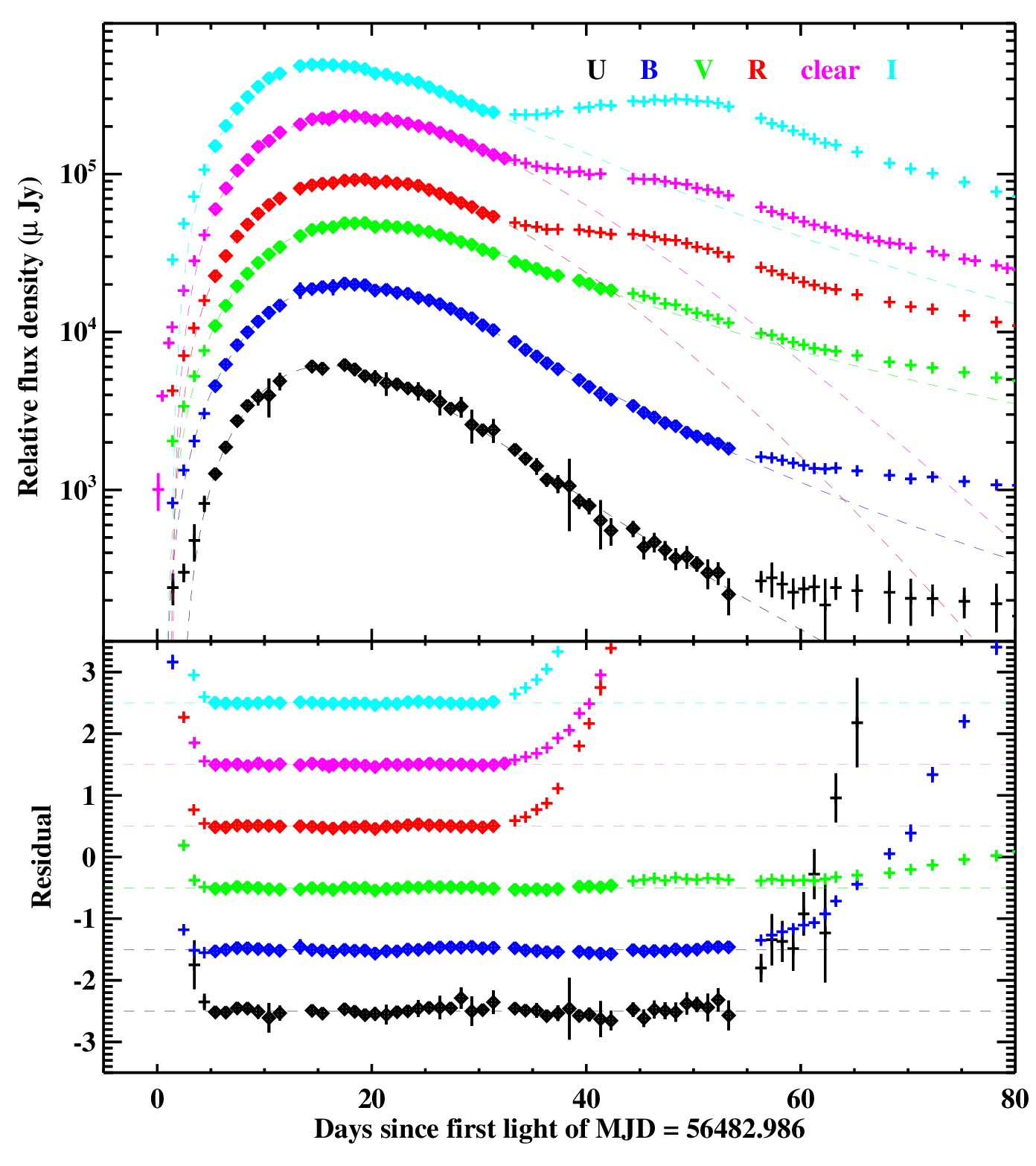}
\caption{Multiband light-curve fitting of SN~2013dy using Equation~\ref{eq_lvtbkn2}
         for estimating $t_0$. The value of $\alpha_r$ is fixed to be 1.90 and $t_b$ 
        is fixed to be 23.0 during the fitting procedure.
        Diamond-shaped data points are included in the fitting
         while cross-shaped ones are excluded.}
\label{13dy_IaBkns_fitting_case53}
\end{figure}

Next, in order to simulate the different discovery times of most SNe, for each of the above
9 well-observed SNe we gradually include data points earlier than $t_{\rm pm-1}{-0}$ until all
data are included. Each step gives a corresponding $t_{\rm pm-1}{X}$ value ($X$ varies from 0 to $-7$). We then use Equation~\ref{eq_lvtbkn2} to fit $t_0$, and we compare the
estimated value with the ``known $t_0$.'' Again, we use SN~2013dy as an example to demonstrate this procedure, which is very similar for all of our 9 SNe in each band.
Figure~\ref{13dy_IaBkns_fitting_case53} illustrates the case of SN~2013dy with $t_{\rm pm-1}{-3}$.
Note the difference with Figure~\ref{13dy_IaBkns_fitting_case1}; now
we include only data after $t_{\rm pm-1}{-3}$ in order to simulate a
late discovery. The data before $t_{\rm pm-1}{-3}$ are shown as crosses, which means they are
not included. As mentioned above, for all fitting to estimate $t_0$, we fix
$\alpha_r$ to be 1.90 and $t_b$ to be 23.0.

\begin{figure}
\centering
\includegraphics[width=.5\textwidth]{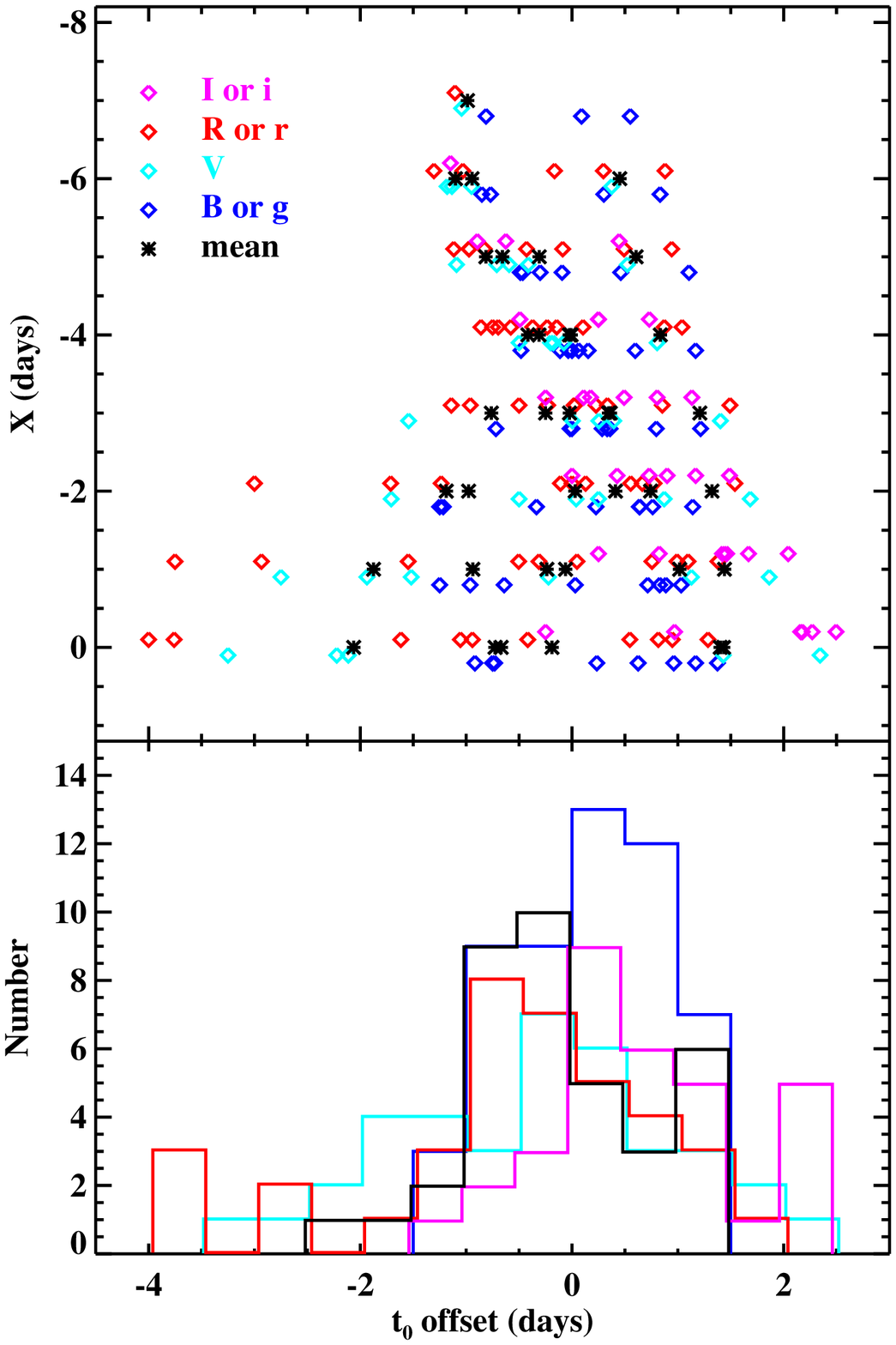}
\caption{Difference of first-light time $t_0$ estimated from Equation~\ref{eq_lvtbkn2} fitting
         with the real $t_0$. Upper panel shows the cases for each SN and each filter with
	 different $t_{\rm pm-1}{X}$, arbitrarily shifted along the ordinate 
         for clarity.
         The bottom panel shows the histogram of $t_0$ offsets for each 
         filter, also arbitrarily shifted along the abscissa for clarity.
         In general, the smaller the $X$ in $t_{\rm pm-1}{X}$ (with earlier data),
	 the smaller the offset. The $B$ (or $g$) band has the smallest offset, and appears
	 to not change with $X$, with a 1$\sigma$ systematic error of $\pm 0.7$ days.}
\label{t0_compare_error}
\end{figure}

After applying the $t_0$ fit to each SN for each band, and for each case with different
$t_{\rm pm-1}{X}$ ($X$ varies from 0 to $-7$), we compare the estimated $t_0$ with the ``known $t_0$.''
The result is shown in Figure~\ref{t0_compare_error}, where the upper panel shows the
cases for each SN and each filter with different $t_{\rm pm-1}{X}$. For a certain $t_{\rm pm-1}{X}$,
if there is more than one filter observed for a SN (as shown in Table 1, except for the
three {\it Kepler} SNe that were only observed with one broadband filter), a mean value of $t_0$ from all
filters is also calculated. The bottom panel shows the histogram of the $t_0$ offsets
compared with the ``known $t_0$''; each filter is color coded.

Three important and clear conclusions can be drawn from Figure~\ref{t0_compare_error}. First,
the offset between the fitting $t_0$ and the ``known $t_0$'' becomes smaller when earlier data are 
included --- namely, the
smaller the $X$ value in $t_{\rm pm-1}{X}$, the smaller the offset. But once $X$ reaches $-3$, then even earlier
data do not help much for $t_0$ estimation. Second, $B$ interestingly
has the smallest offset, and no matter when starting with $t_{\rm pm-1}{X}$, 
the offset is always within $\pm1.5$ days; we estimate a 1$\sigma$ 
systematic error of $\pm 0.7$ days
by adopting only the $B$-band estimate of the $t_0$ offset.
Third, the mean offset with multiple filters is similar to that of the $B$ band; thus, for
$t_0$ estimation, one should use either the mean value or the value estimated from $B$
alone. Technically, it is not surprising to see why the $B$ band gives the
best estimate of $t_0$; in other bands, the peak is usually broader than that of $B$,
making $B$ better for estimating $t_0$. Because of this, in the following
we only adopt the $t_0$ estimate from $B$, even if the SN was
observed with multiple filters. We will also add a systematic error of $\pm 0.7$ days for
the $t_0$ estimated from $B$-band fitting.

In addition, Figure~\ref{t0_compare_error} shows that the fixed value of $\alpha_r = 1.90$ is
appropriate, giving a mean $t_0$ offset of $0.1 \pm 0.7$ days for the $B$ band, very close to 0
as expected.
We further performed two other sets of fitting with the $\alpha_r$ value fixed
to be 1.50 and 2.30 (respectively), and found $t_0$ offset by $1.2 \pm 0.8$ and
$-0.8 \pm 0.7$ for $B$, confirming that the fixed value of $\alpha_r = 1.90$ is appropriate.

\subsection{A Larger Well-Observed Sample for $t_0$ Estimation}

From the above analysis, we have shown that Equation~\ref{eq_lvtbkn2} provides a
practical way to estimate $t_0$ by fixing $\alpha_r$ (to be 1.90) and $t_b$ (to be 23.0) during fitting,
and using only $B$ (or $g$) data. Such an estimate of $t_0$
gives an additional systematic error of $\pm0.7$ days. We can now apply this method to
a larger sample of objects that were typically discovered around $t_{\rm pm-1}{-0}$, or slightly earlier,
but still with good-coverage observations after discovery.

Ganeshalingam et al. (2010) published a sample of 165 SNe~Ia from the Lick Observatory Supernova Search
(LOSS: Filippenko et al. 2001; Leaman et al. 2011) database. From this sample,
we selected 44 SNe~Ia that are suitable for our purpose. As above, all satisfy the
criterion that the first observation is earlier 
than the time that the SN brightness reaches one magnitude fainter than the peak
($t_{\rm pm-1}{-0}$). (Of course, even earlier observations are preferred.)
This lower limit criterion allowed us to choose 44 of the 165 LOSS SNe~Ia.
In addition, 6 SNe~Ia were selected from the Harvard-Smithsonian Center for Astrophysics
Data Release 3 (CfA3; Hicken et al. 2009), as well as 6 SNe~Ia from the Carnegie Supernova
Project (CSP; Contreras et al. 2010),
making the final sample 56 SNe~Ia, as listed in Table~2.

\subsection{Rise-Time Estimation}

We applied the fitting to the above 56 SNe~Ia using Equation~\ref{eq_lvtbkn2} only with $B$ data
(no $g$-band data were collected for these 56 SNe~Ia),
and fixing $\alpha_r$ to be 1.90 and $t_b$ to be 23.0.
The resulting $t_0$ estimates are listed in Table~2.

Once the first-light time $t_0$ is estimated, it is easy to calculate the rise time
as long as the time of peak brightness can also be determined. 
Technically, it is much easier to estimate the peak time than the first-light time $t_0$.
Although the above fitting method can also give the peak time,
it is more accurate and straightforward to estimate the peak time directly by fitting the
light curve around peak brightness with a low-order polynomial. The time of peak
$B$-band brightness is given in Table~2, along with
the decline rate $\Delta m_{15}(B)$. The rise time is simply the time duration
from first-light time to the time of peak brightness, corrected for time dilation through division by $1 + z$; see Table~2 for the final results.


\section{Discussion}\label{s:discussion}

Unlike traditional light-curve fitting, which usually compares with templates,
our estimation of the first-light time $t_0$ and rise time $t_r$ are model independent;
they are direct measurements of the observed data.

\begin{figure*}
\centering
\includegraphics[width=.49\textwidth]{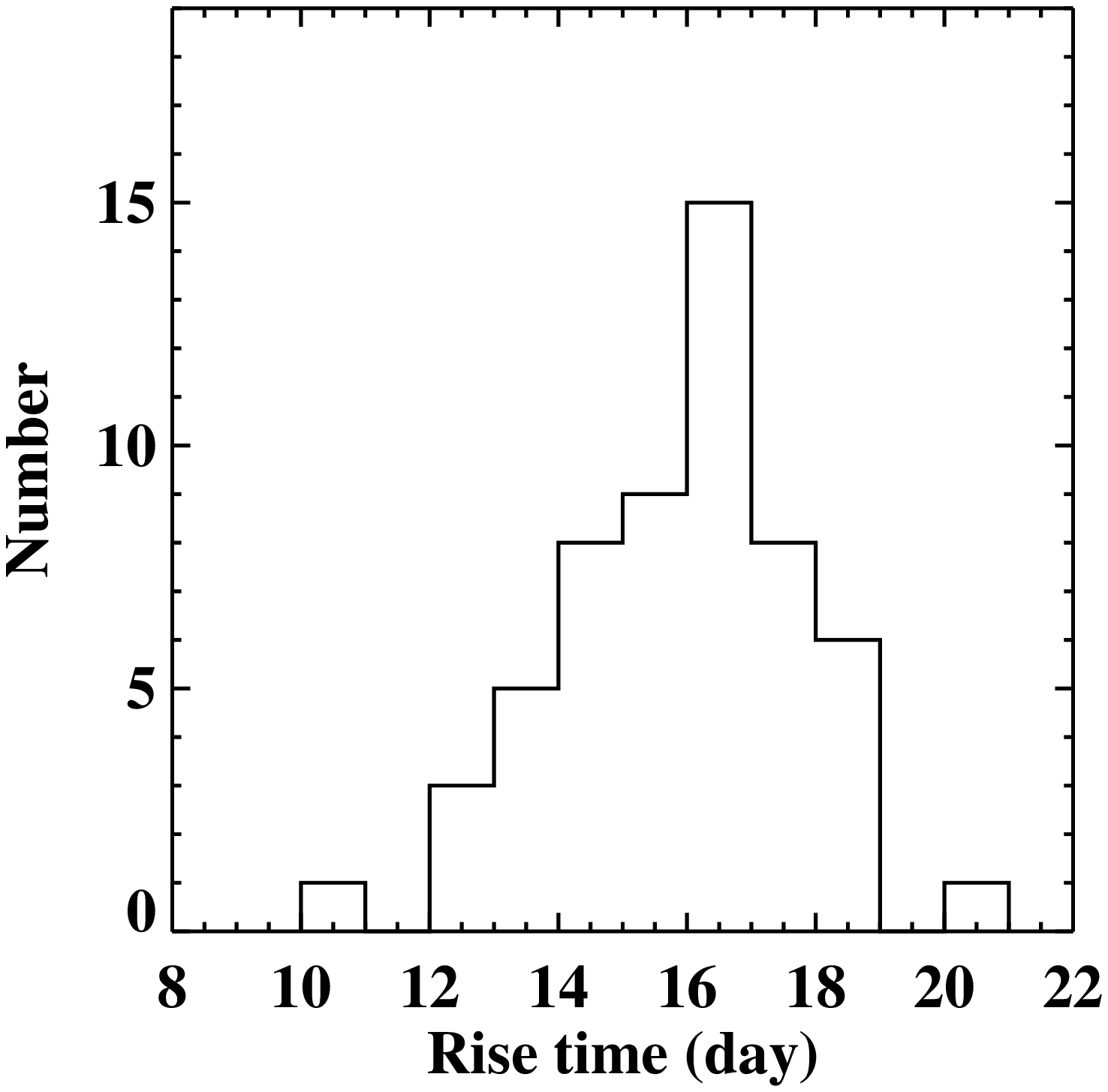}
\includegraphics[width=.49\textwidth]{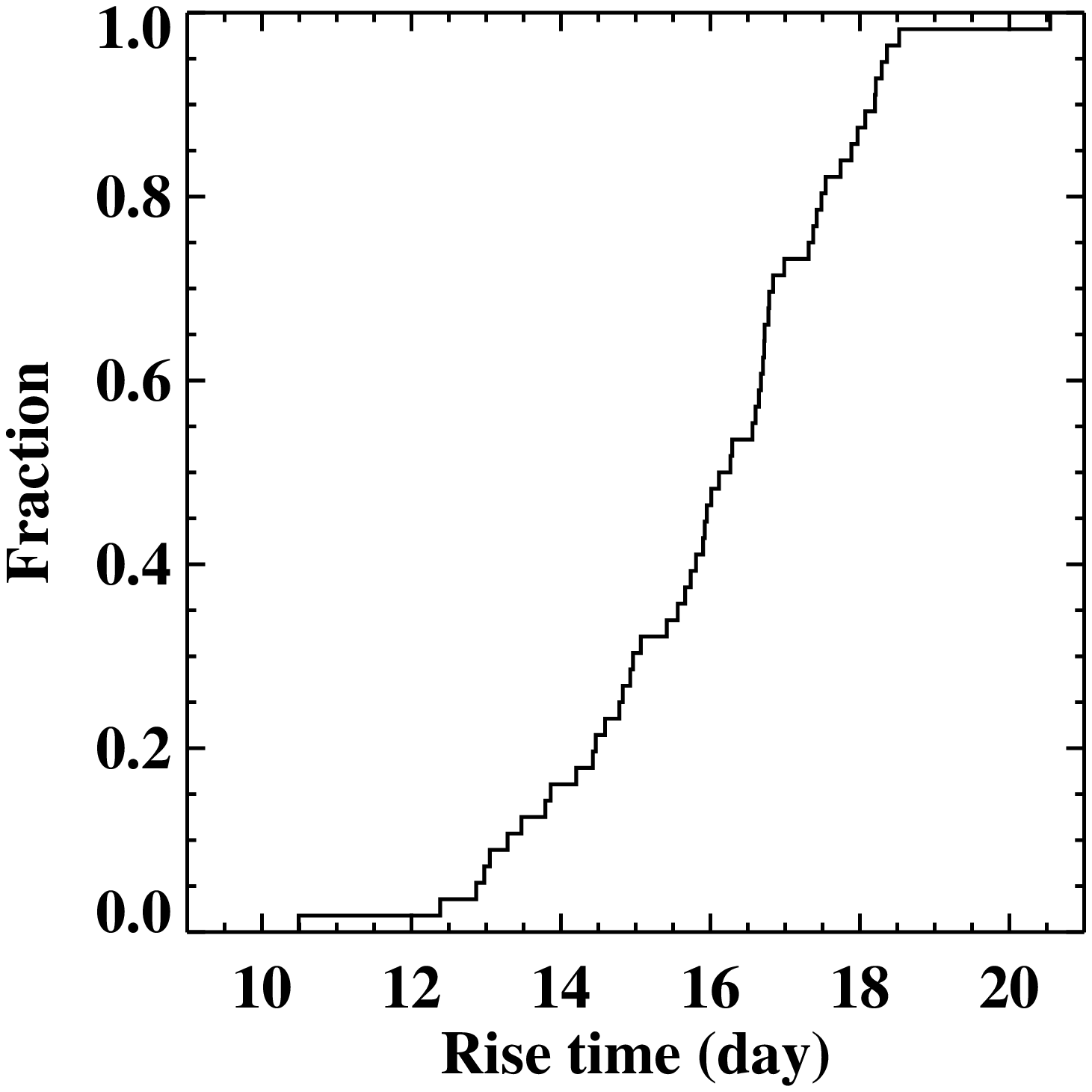}
\caption{Histogram (left panel) and cumulative (right panel) distribution of rise times $t_r$ from
         the 56 SNe~Ia in our sample. The mean rise time is 16.0 days.}
\label{tr_hist_cumulative}
\end{figure*}

Figure~\ref{tr_hist_cumulative} shows the histogram (left) and the cumulative (right panel)
distribution of $t_r$ of the 56 SNe~Ia in our sample. The rise time varies from
10.5 days to 20.5 days with quite a wide range and concentrated around 16.0 days. The shortest
rise time (10.5 days) is for SN 2003Y, a SN 1991bg-like (e.g., Filippenko et al. 1992a)
 subluminous SN~Ia, while the longest (20.5 days) is
for SN 2005M, a SN 1991T-like (e.g., Filippenko et al. 1992b) overluminous SN~Ia.
The remaining SNe have rise times between 12.0 and 19.0 days,
with a mean value of 16.0 days, slightly longer than two weeks.
Note that our estimate is slightly shorter than the mean rise time of 18.98 days
derived by Firth et al. (2015), probably because their fitting model differs from ours.

\begin{figure*}
\centering
\includegraphics[width=.49\textwidth]{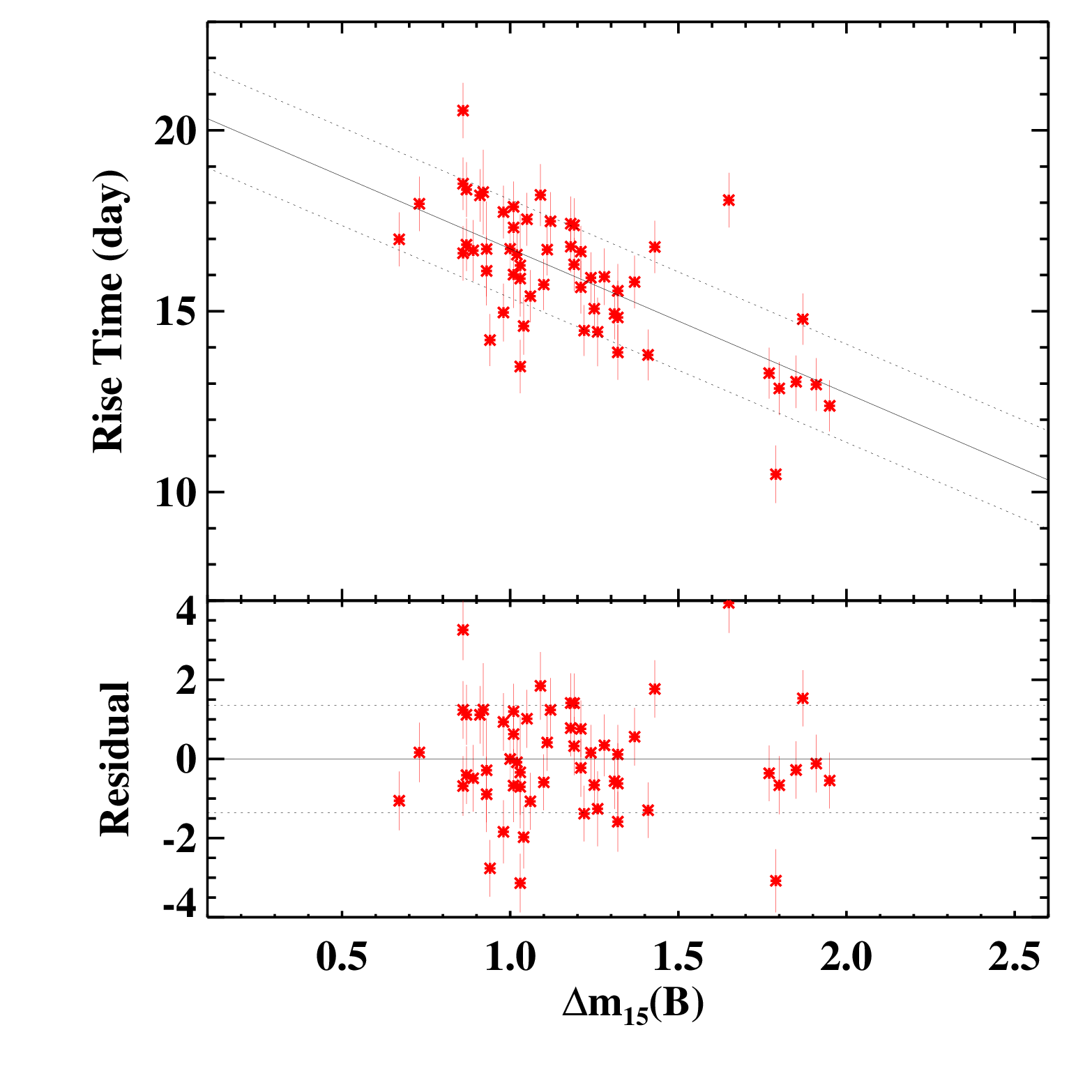}
\includegraphics[width=.49\textwidth]{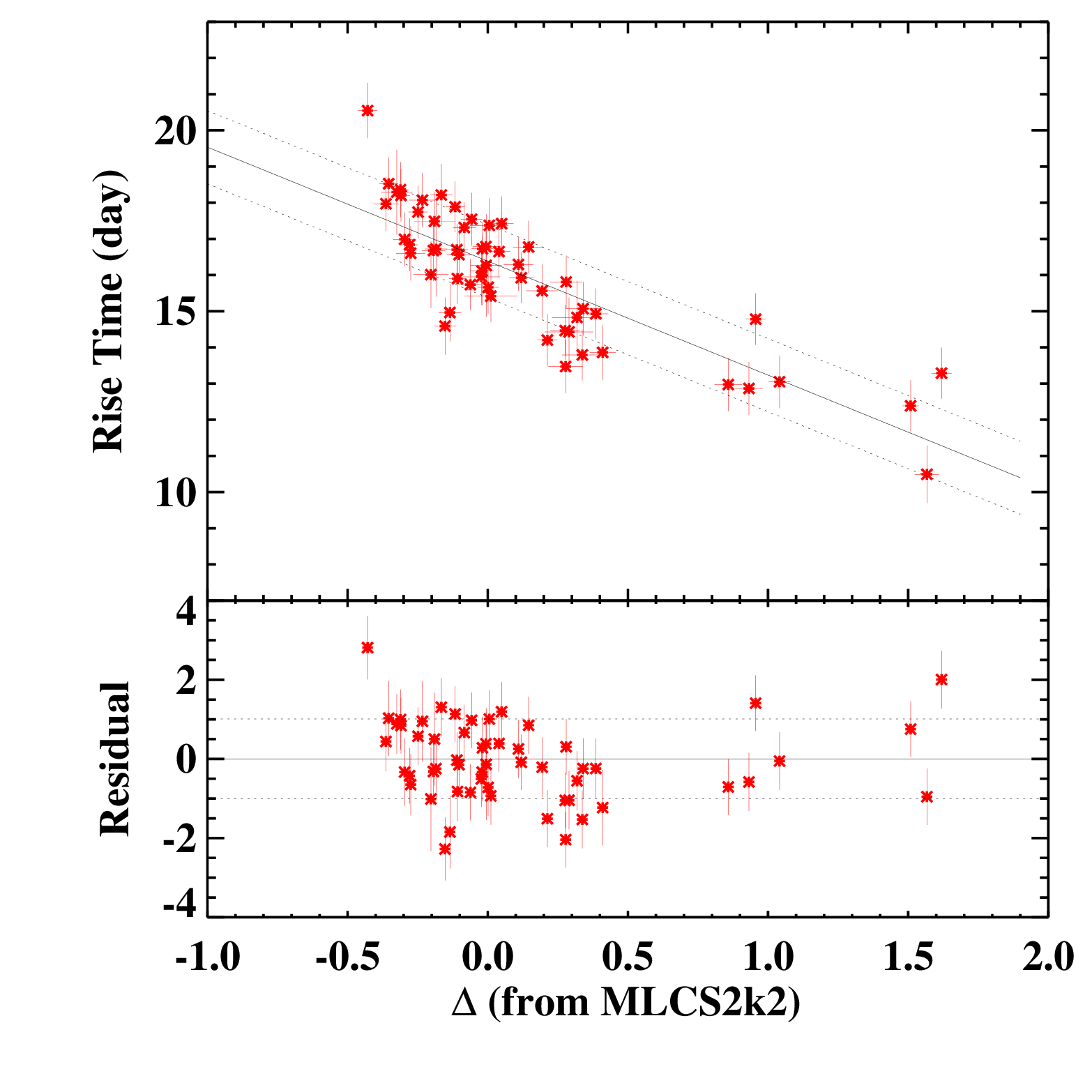}
\caption{Rise time $t_r$ as a function of decline rate $\Delta m_{15}(B)$ (left panel)
         and $\Delta$ from MLCS2k2 fitting (right panel), with a smaller scatter
	 for the latter one. The bottom panels show the residuals from the 
         best-fit relation. Dotted lines mark the 1$\sigma$ scatter.
         }
\label{tr_compare_delta_dm15}
\end{figure*}

In Figure~\ref{tr_compare_delta_dm15}, we compare the rise time $t_r$ with two other important
parameters for SNe~Ia.
At left, we plot $t_r$ as a function of decline rate $\Delta m_{15}(B)$.
There is an apparent correlation between the two parameters: the larger the rise time, the smaller the decline rate,
and our best fit gives $t_r = -4.0 \times \Delta m_{15}(B) + 20.7$.
However, this relation has a 1$\sigma$ scatter of 1.3 days,
estimated from the residuals of the whole sample, as shown in the 
bottom-left panel of Figure~\ref{tr_compare_delta_dm15}.
This large scatter indicates that a single decline-rate parameter might not
be enough to characterize the SN~Ia light curve.
In the right-hand panel, we plot $t_r$ as a function of
$\Delta$ from MLCS2k2 fitting (Jha et al. 2007), known as the stretch parameter. 
There is also a clear correlation: larger rise times have smaller values of $\Delta$,
and our best fit gives $t_r = -3.2 \times \Delta + 16.4$.
This relation is substantially tighter than the previous one, with
a 1$\sigma$ scatter of only 1.0 days.

It is not surprising that the scatter between the rise time with $\Delta$
is smaller than that with $\Delta m_{15}(B)$, because the stretch parameter
$\Delta$ considers both the rising and declining portions of the light curve in the
MLCS2k2 fitting (Jha et al. 2007).
Our result confirms the Hayden et al. (2010) and Ganeshalingam et al. (2010)
conclusion that a single parameter is not enough to characterise the light curve.
However, here we focus on studying the rising part of the light curve, at which
time the emission from a SN~Ia better approximates a blackbody as was assumed for
deriving the fitting function Equation~\ref{eq_lvtbkn2}; the SN is optically thicker
at early times than during the decay.
Although the rising and decaying parts are correlated, our results indicate that the scatter is large,
so for some purposes it is better to distinguish the two sections.

Interestingly, but not surprisingly, most of the short-rise-time objects are subluminous
SNe~Ia (similar to the subtype of SN 1991bg-like SNe~Ia), while those with longer rise times are
overluminous SNe~Ia (similar to the subtype of SN 1991T-like and SN 1999aa-like SNe~Ia, see
Li et al. (2001)). 

In our sample, among the six objects with $t_r < 13.5$ days, 
three [SN~2003Y (Matheson et al. 2003), SN~2005ke (Patat et al. 2005),
and SN~1999by (Howell et al. 2001)] are spectroscopically classified as 
SN~1991bg-like.
SN~2002dl is also spectroscopically similar (Matheson et al. 2002) to the 
SN~1991bg-like SN~1999by.
For one object (SN~2000dr), a spectrum from the UC Berkeley Supernova Database (UCB SNDB; Silverman et al. 2012a)
\footnote{The SNDB was updated in 2015 and is available online at http://heracles.astro.berkeley.edu/sndb/},
shows that it best matches several normal SNe~Ia about one week after
maximum brightness based on the
SN IDentication code (SNID; Blondin \& Tonry 2007), but it also matches 
the SN~1991bg-like SN~2007ba at +5 days;
thus, SN~2000dr is also a possible SN~1991bg-like object.
Spectra from UCB SNDB show that the final object (SN~2007ci)
is likely a normal SN~Ia.
According to the five possible SN~1991bg-like SNe (including SN~2000dr), 
a good criterion for
distinguishing SN~1991bg-like objects is $t_r < 13.5$ days
if only using the rise time, but with one exception: 
SN~2007ci has a rise time of 13.0 days but is a normal SN~Ia.

Similarly, among the 8 objects with $t_r > 18.0$ days,
five of them [SN~1999dq (Jha et al. 1999), SN~2003fa (see spectra from 
the UCB SNDB), SN~2005M (Thomas 2005), SN~1999gp (Jha \& Berlind 2000), 
and SN~2002eb (see spectra from UCB the SNDB)] are classified as 
SN~1991T-like or SN~1999aa-like objects.
One of them (SN~2005hk) belongs to the SN~Iax subclass (McCully et al. 2014).
The other two [SN~2006ax (see spectra from Blondin et al. 2012)
and SN~2006gr (see spectra from the UCB SNDB)] best match
several normal SNe~Ia but also a few SN~1999aa-like SNe; thus,
SN~2006ax and SN~2006gr are also possibly SN~1999aa-like objects.
According to the seven possible SN~1991T-like or SN~1999aa-like 
SNe (including SN~2006ax and SN~2006gr), a good criterion for
distinguishing SN~1991T-like and SN~1999aa-like objects is 
$t_r > 18.0$ days if only using the rise time, 
but with one exception: SN~2005hk has a rise time of 18.1 days 
but is a SN~Iax.

The above analysis shows that the rise time is an additional 
parameter helpful for classifying SN~Ia subtypes.


\section{Conclusions}\label{s:conclusions}

We have adopted an empirical method to fit the optical light curves of SNe~Ia
which is useful for estimating their first-light time and rise time.
Our method differs from the usual template-fitting method for SN~Ia light curves;
we give direct measurements of the first-light time and rise time.
From a sample of 56 well-observed SNe~Ia, we find that the rise time ranges
from 10.5 days to 20.5 days, with a mean rise time of 16.0 days. Our results confirm that
the rise time is generally correlated with the decline rate $\Delta m_{15}(B)$ and
stretch parameter $\Delta$, but with large scatter, especially when using $\Delta m_{15}(B)$.
We also show that the rise time is useful for SN~Ia subtype classification.

\bigskip
\medskip


%
We thank Isaac Shivvers and Melissa L. Graham for useful discussions and suggestions,
as well as the staffs of the observatories where data were obtained. 
We also thank Brad E. Tucker for providing data for the three {\it Kepler} SNe~Ia published by Olling
et al. (2015).
A.V.F.'s supernova group at UC Berkeley is grateful for financial
assistance from NSF grant AST-1211916, the TABASGO Foundation,
the Christopher R. Redlich Fund, and the Miller Institute for Basic
Research in Science (U.C. Berkeley). The work of A.V.F. was completed
in part at the Aspen Center for Physics, which is supported by NSF
grant PHY-1607611; he thanks the Center for its hospitality during
the neutron stars workshop in June and July 2017.







%
%
%
%

\clearpage
\begin{deluxetable}{cccccccc}
 \tabcolsep 0.4mm
 \tablewidth{0pt}
 \tablecaption{The 9 Extremely Well-Observed Type~Ia Supernovae}
  \tablehead{\colhead{SN} & \colhead{known $t_0$(MJD)$^a$} & \colhead{filter} & \colhead{$t_{b}$} & \colhead{$\alpha_{r}$} & \colhead{$\alpha_{d}$} & \colhead{$s$} & \colhead{$\chi^2$/dof}}
\startdata
ASASSN-14lp  &  56998.39     &   $V$  & 23.2$\pm$0.7   & 1.48$\pm$0.13   & 2.25$\pm$0.10   & 1.38$\pm$0.14 & 109.43  / 20 = 5.47   \\
ASASSN-14lp  &               &   $u$  & 25.0$\pm$0.8   & 2.81$\pm$0.08   & 5.37$\pm$0.18   & 0.36$\pm$0.03 & 350.62  / 29 = 12.09  \\
ASASSN-14lp  &               &   $g$  & 33.5$\pm$1.3   & 1.88$\pm$0.04   & 4.37$\pm$0.17   & 0.44$\pm$0.03 & 366.15  / 36 = 10.17  \\
ASASSN-14lp  &               &   $r$  & 25.7$\pm$1.8   & 1.65$\pm$0.03   & 2.80$\pm$0.24   & 0.86$\pm$0.13 & 301.56  / 22 = 13.70  \\
ASASSN-14lp  &               &   $i$  & 15.0$\pm$0.4   & 1.73$\pm$0.03   & 1.82$\pm$0.06   & 1.58$\pm$0.14 & 156.04  / 19 = 8.21   \\
  iPTF13ebh  &  56607.85     &   $B$  & 18.7$\pm$0.2   & 1.91$\pm$0.04   & 3.58$\pm$0.08   & 1.24$\pm$0.06 & 49.52   / 16 = 3.09   \\
  iPTF13ebh  &               &   $V$  & 20.8$\pm$0.5   & 1.90$\pm$0.03   & 2.93$\pm$0.13   & 1.11$\pm$0.10 & 100.95  / 16 = 6.30   \\
  iPTF13ebh  &               &   $u$  & 15.8$\pm$0.3   & 2.53$\pm$0.11   & 3.90$\pm$0.16   & 0.94$\pm$0.11 & 32.83   / 13 = 2.52   \\
  iPTF13ebh  &               &   $g$  & 21.7$\pm$0.4   & 1.99$\pm$0.03   & 4.07$\pm$0.13   & 0.84$\pm$0.05 & 176.37  / 16 = 11.02  \\
  iPTF13ebh  &               &   $r$  & 14.9$\pm$0.1   & 1.80$\pm$0.03   & 1.58$\pm$0.02   & 3.44$\pm$0.19 & 94.93   / 16 = 5.93   \\
  iPTF13ebh  &               &   $i$  & 12.8$\pm$0.2   & 1.87$\pm$0.05   & 1.59$\pm$0.05   & 3.16$\pm$0.39 & 58.09   / 11 = 5.28   \\
     2011fe  &  55796.687    &   $B$  & 27.0$\pm$0.7   & 2.52$\pm$0.16   & 4.75$\pm$0.15   & 0.53$\pm$0.03 & 315.26  / 129= 2.44   \\
     2011fe  &               &   $V$  & 19.5$\pm$0.1   & 2.18$\pm$0.04   & 2.33$\pm$0.02   & 1.40$\pm$0.04 & 548.44  / 126= 4.35   \\
     2011fe  &               &   $R$  & 20.7$\pm$0.4   & 2.17$\pm$0.03   & 2.70$\pm$0.08   & 1.12$\pm$0.07 & 444.17  / 107= 4.15   \\
     2011fe  &               &   $I$  & 15.0$\pm$0.3   & 2.28$\pm$0.04   & 2.14$\pm$0.06   & 1.69$\pm$0.14 & 475.75  / 105= 4.53   \\
     2011fe  &               &   $g$  & 24.0$\pm$0.3   & 2.21$\pm$0.03   & 3.21$\pm$0.05   & 0.80$\pm$0.03 & 2892.49 / 504= 5.73   \\
     2009ig  &  55062.910    &   $B$  & 30.8$\pm$3.6   & 1.66$\pm$0.10   & 4.25$\pm$0.61   & 0.56$\pm$0.15 & 83.81   / 28 = 2.99   \\
     2009ig  &               &   $V$  & 18.7$\pm$0.4   & 1.51$\pm$0.11   & 1.65$\pm$0.05   & 2.41$\pm$0.24 & 153.90  / 28 = 5.49   \\
     2009ig  &               &   $R$  & 26.9$\pm$3.7   & 1.68$\pm$0.05   & 2.99$\pm$0.47   & 0.74$\pm$0.20 & 174.66  / 24 = 7.27   \\
     2009ig  &               &   $I$  & 20.0$\pm$3.1   & 1.86$\pm$0.10   & 2.53$\pm$0.42   & 0.83$\pm$0.29 & 43.85   / 22 = 1.99   \\
     2013dy  &  56482.986    &   $B$  & 23.4$\pm$0.6   & 1.87$\pm$0.07   & 3.09$\pm$0.11   & 0.97$\pm$0.11 & 35.30   / 42 = 0.84   \\
     2013dy  &               &   $V$  & 19.7$\pm$0.3   & 1.77$\pm$0.04   & 1.99$\pm$0.04   & 1.70$\pm$0.14 & 44.92   / 42 = 1.06   \\
     2013dy  &               &   $R$  & 25.1$\pm$3.0   & 1.85$\pm$0.06   & 2.97$\pm$0.45   & 0.82$\pm$0.23 & 43.97   / 26 = 1.69   \\
     2013dy  &               &   $I$  & 18.5$\pm$1.8   & 1.85$\pm$0.10   & 2.35$\pm$0.32   & 1.15$\pm$0.35 & 8.88    / 23 = 0.38   \\
     2013dy  &               &   $U$  & 22.7$\pm$1.7   & 2.18$\pm$0.17   & 3.82$\pm$0.36   & 0.71$\pm$0.18 & 15.51   / 42 = 0.36   \\
     2012cg  &  56063.950    &   $B$  & 28.1$\pm$3.7   & 2.06$\pm$0.10   & 4.17$\pm$0.67   & 0.58$\pm$0.17 & 11.44   / 28 = 0.40   \\
     2012cg  &               &   $V$  & 20.9$\pm$0.8   & 1.76$\pm$0.04   & 2.10$\pm$0.11   & 1.63$\pm$0.24 & 20.20   / 28 = 0.72   \\
     2012cg  &               &   $R$  & 22.5$\pm$2.1   & 1.83$\pm$0.07   & 2.54$\pm$0.32   & 1.13$\pm$0.30 & 37.72   / 21 = 1.79   \\
     2012cg  &               &   $I$  & 16.4$\pm$0.9   & 1.85$\pm$0.08   & 1.95$\pm$0.16   & 1.66$\pm$0.39 & 14.67   / 19 = 0.77   \\
  KSN-2011b  &  995.710 $^b$ & broad  & 23.1$\pm$0.3   & 2.38$\pm$0.02   & 3.33$\pm$0.06   & 0.71$\pm$0.03 & 2185.09 / 57 = 38.33  \\
  KSN-2011c  &  1075.914$^b$ & broad  & 22.7$\pm$0.4   & 1.77$\pm$0.05   & 2.63$\pm$0.12   & 1.97$\pm$0.23 & 303.76  / 60 = 5.06   \\
  KSN-2012a  &  1328.466$^b$ & broad  & 22.0$\pm$0.7   & 2.01$\pm$0.04   & 3.22$\pm$0.13   & 0.70$\pm$0.06 & 1153.31 / 53 = 21.76  \\
\enddata
\tablenotetext{a}{References are given in Section \ref{s:data_analysis1}, and data within 3.0 days after $t_0$ are not included in the fitting.}
\tablenotetext{b}{KJD = MJD$-$54832.5. Its broadband filter is plotted as the $R$ band.}
\end{deluxetable}

\clearpage
\begin{deluxetable}{llccccccc}
 \tabcolsep 0.4mm
 \tablewidth{0pt}
 \tabletypesize{\scriptsize}
 \tablecaption{The 56 Well-Observed Type~Ia Supernovae}
  \tablehead{\colhead{SN} & \colhead{subtype} & \colhead{$z$} & \colhead{$t_{0,B}$} & \colhead{$t_{0,B}$ err$^a$} & \colhead{$t_{p,B}$} & \colhead{$t_{r,B}$} & \colhead{$\Delta m_{15}(B)$} & \colhead{$\Delta$}}
\startdata
\multicolumn{9}{c} {From LOSS}   \\
\hline
\\
  1998dh  &  Ia-norm  &     0.0077  &         2451013.6  &     0.7  &         2451029.4  &    15.7  &    1.21  &    0.0015  \\
  1998dm  &  Ia-norm  &     0.0055  &         2451043.1  &     0.7  &         2451060.9  &    17.7  &    0.98  &   -0.2488  \\
  1999by  &  Ia-91bg  &     0.0027  &         2451296.1  &     0.7  &         2451308.5  &    12.4  &    1.95  &    1.5087  \\
  1999cp  &  Ia-norm  &     0.0103  &         2451346.3  &     0.7  &         2451363.8  &    17.3  &    1.01  &   -0.0837  \\
  1999dq  &  Ia-99aa  &     0.0137  &         2451417.8  &     0.8  &         2451436.4  &    18.4  &    0.87  &   -0.3112  \\
  1999gp  &  Ia-norm  &     0.0260  &         2451531.1  &     0.8  &         2451549.6  &    18.0  &    0.73  &   -0.3633  \\
  2000cx  &   Ia-pec  &     0.0070  &         2451738.0  &     0.7  &         2451752.3  &    14.2  &    0.94  &    0.2128  \\
  2000dn  &  Ia-norm  &     0.0308  &         2451807.7  &     1.4  &         2451824.5  &    16.3  &    1.03  &   -0.0058  \\
  2000dr  &  Ia-norm  &     0.0178  &         2451821.0  &     0.7  &         2451834.1  &    12.9  &    1.80  &    0.9319  \\
  2000fa  &  Ia-norm  &     0.0218  &         2451874.7  &     0.8  &         2451891.7  &    16.7  &    0.89  &   -0.1939  \\
  2001en  &  Ia-norm  &     0.0153  &         2452176.2  &     0.7  &         2452192.8  &    16.3  &    1.19  &    0.1095  \\
  2001ep  &  Ia-norm  &     0.0129  &         2452183.0  &     0.7  &         2452200.0  &    16.8  &    1.43  &    0.1458  \\
  2002bo  &  Ia-norm  &     0.0053  &         2452340.9  &     0.7  &         2452356.7  &    15.7  &    1.10  &   -0.0620  \\
  2002cr  &  Ia-norm  &     0.0103  &         2452391.9  &     0.7  &         2452408.7  &    16.6  &    1.21  &    0.0403  \\
  2002dj  &  Ia-norm  &     0.0104  &         2452435.0  &     0.7  &         2452450.6  &    15.4  &    1.06  &    0.0109  \\
  2002dl  &   Ia-pec  &     0.0152  &         2452439.2  &     0.7  &         2452452.5  &    13.0  &    1.85  &    1.0411  \\
  2002eb  &  Ia-norm  &     0.0265  &         2452475.8  &     0.7  &         2452494.5  &    18.2  &    0.91  &   -0.3094  \\
  2002er  &  Ia-norm  &     0.0090  &         2452508.4  &     0.7  &         2452524.5  &    15.9  &    1.24  &    0.1195  \\
  2002fk  &  Ia-norm  &     0.0070  &         2452530.0  &     0.7  &         2452548.0  &    17.9  &    1.01  &   -0.1163  \\
  2002ha  &  Ia-norm  &     0.0132  &         2452566.2  &     0.7  &         2452581.3  &    14.9  &    1.31  &    0.3856  \\
  2002he  &  Ia-norm  &     0.0248  &         2452571.6  &     0.8  &         2452585.9  &    13.9  &    1.32  &    0.4099  \\
  2003cg  &  Ia-norm  &     0.0053  &         2452713.2  &     1.0  &         2452729.4  &    16.1  &    0.93  &   -0.0201  \\
  2003fa  &  Ia-99aa  &     0.0391  &         2452788.0  &     0.7  &         2452807.3  &    18.5  &    0.86  &   -0.3534  \\
  2003gn  &  Ia-norm  &     0.0333  &         2452837.8  &     0.9  &         2452852.7  &    14.4  &    1.26  &    0.2899  \\
  2003gt  &  Ia-norm  &     0.0150  &         2452845.0  &     0.7  &         2452862.0  &    16.7  &    1.00  &   -0.0191  \\
   2003W  &  Ia-norm  &     0.0211  &         2452664.0  &     0.8  &         2452678.9  &    14.6  &    1.04  &   -0.1525  \\
   2003Y  &  Ia-91bg  &     0.0173  &         2452665.9  &     0.8  &         2452676.6  &    10.5  &    1.79  &    1.5668  \\
  2004at  &  Ia-norm  &     0.0240  &         2453075.2  &     0.7  &         2453092.1  &    16.6  &    1.02  &   -0.1025  \\
  2004dt  &  Ia-norm  &     0.0185  &         2453223.3  &     0.7  &         2453240.3  &    16.7  &    1.11  &   -0.1102  \\
  2004ef  &  Ia-norm  &     0.0298  &         2453250.0  &     0.7  &         2453264.2  &    13.8  &    1.41  &    0.3373  \\
  2004eo  &  Ia-norm  &     0.0148  &         2453262.4  &     0.7  &         2453278.5  &    15.8  &    1.37  &    0.2798  \\
  2005cf  &  Ia-norm  &     0.0070  &         2453517.7  &     0.7  &         2453533.7  &    15.9  &    1.03  &   -0.1080  \\
  2005de  &  Ia-norm  &     0.0149  &         2453581.1  &     0.7  &         2453598.8  &    17.4  &    1.18  &    0.0495  \\
  2005ki  &  Ia-norm  &     0.0203  &         2453690.0  &     0.7  &         2453705.4  &    15.1  &    1.25  &    0.3400  \\
   2005M  &   Ia-91T  &     0.0230  &         2453384.7  &     0.8  &         2453405.8  &    20.5  &    0.86  &   -0.4287  \\
  2006cp  &  Ia-norm  &     0.0233  &         2453879.6  &     0.8  &         2453897.5  &    17.5  &    1.12  &   -0.1903  \\
  2006gr  &  Ia-norm  &     0.0335  &         2453993.5  &     1.2  &         2454012.4  &    18.3  &    0.92  &   -0.3245  \\
  2006le  &  Ia-norm  &     0.0172  &         2454030.8  &     0.8  &         2454047.7  &    16.6  &    0.86  &   -0.2751  \\
   2006X  &  Ia-norm  &     0.0064  &         2453770.4  &     0.8  &         2453786.5  &    16.0  &    1.28  &   -0.0228  \\
  2007af  &  Ia-norm  &     0.0062  &         2454157.5  &     0.7  &         2454174.4  &    16.8  &    1.18  &   -0.0064  \\
  2007le  &  Ia-norm  &     0.0067  &         2454383.7  &     0.8  &         2454398.8  &    15.0  &    0.98  &   -0.1349  \\
  2007qe  &  Ia-norm  &     0.0244  &         2454412.6  &     0.9  &         2454429.0  &    16.0  &    1.01  &   -0.2030  \\
  2008bf  &  Ia-norm  &     0.0251  &         2454537.6  &     1.3  &         2454554.7  &    16.7  &    0.93  &   -0.1849  \\
  2008ec  &  Ia-norm  &     0.0149  &         2454658.2  &     0.7  &         2454674.0  &    15.6  &    1.32  &    0.1941  \\
\hline
\\
\multicolumn{9}{c} {From CfA3}   \\
\hline
\\
   2001V  &  Ia-norm  &     0.0162  &         51955.183  &     0.7  &         51972.446  &    17.0  &    0.67  &   -0.2967  \\
  2005hk  &      Iax  &     0.0118  &         53666.101  &     0.8  &         53684.385  &    18.1  &    1.65  &   -0.2331  \\
  2006ax  &  Ia-norm  &     0.0180  &         53808.486  &     0.9  &         53827.026  &    18.2  &    1.09  &   -0.1657  \\
  2006lf  &  Ia-norm  &     0.0130  &         54030.365  &     1.0  &         54045.385  &    14.8  &    1.32  &    0.3188  \\
  2007bd  &  Ia-norm  &     0.0319  &         54191.882  &     0.7  &         54205.783  &    13.5  &    1.03  &    0.2778  \\
  2007ci  &  Ia-norm  &     0.0194  &         54233.021  &     0.7  &         54246.246  &    13.0  &    1.91  &    0.8581  \\
\hline
\\
\multicolumn{9}{c} {From CSP}   \\
\hline
\\
  2005kc  &  Ia-norm  &     0.0137  &         3679.7901  &     0.7  &         3697.4031  &    17.4  &    1.19  &    0.0055  \\
  2005ke  &  Ia-91bg  &     0.0045  &         3684.9211  &     0.7  &         3698.2680  &    13.3  &    1.77  &    1.6190  \\
  2007on  &  Ia-norm  &     0.0062  &         4404.8806  &     0.7  &         4419.7535  &    14.8  &    1.87  &    0.9558  \\
  2008bc  &  Ia-norm  &     0.0157  &         4532.0783  &     0.7  &         4549.1804  &    16.8  &    0.87  &   -0.2790  \\
  2008gp  &  Ia-norm  &     0.0328  &         4760.8915  &     0.7  &         4779.0086  &    17.5  &    1.05  &   -0.0576  \\
  2008hv  &  Ia-norm  &     0.0125  &         4801.8928  &     0.7  &         4816.5393  &    14.5  &    1.22  &    0.2761  \\
\enddata
\tablenotetext{a}{include the 0.7 days systematic error estimated from the fitting method}
\end{deluxetable}

\end{document}